\documentclass[superscriptaddress,onecolumn,reprint]{revtex4-2}
\usepackage{graphicx}
\usepackage{amssymb}
\usepackage{epstopdf}
\usepackage{amsmath,dsfont, commath}
\usepackage{bm}
\usepackage{changes}
\usepackage{physics}
\usepackage{braket}
\usepackage{slashed}

\newcommand{\be}{\begin{equation}}
	\newcommand{\ee}{\end{equation}}
\newcommand{\bea}{\begin{eqnarray}}
	\newcommand{\eea}{\end{eqnarray}}
\newcommand{\ba}{\begin{array}}
	\newcommand{\ea}{\end{array}}

\newcommand{\bl}{\begin{flalign}}
	\newcommand{\enl}{\end{flalign}}

\newcommand{\pa}{\partial}
\newcommand{\mc}[1]{\mathcal{#1}}

\newcommand{\eq}[1]{Eq. \eqref{#1}}

\newcommand{\fig}[1]{Fig. (\ref{#1})}

\newcommand{\half}{\frac{1}{2}}

\renewcommand{\bf}[1]{\mathbf{#1}}
\renewcommand{\Re}{\operatorname{Re}}

\makeatletter
\newsavebox{\@brx}
\newcommand{\llangle}[1][]{\savebox{\@brx}{\(\m@th{#1\langle}\)}%
	\mathopen{\copy\@brx\kern-0.5\wd\@brx\usebox{\@brx}}}
\newcommand{\rrangle}[1][]{\savebox{\@brx}{\(\m@th{#1\rangle}\)}%
	\mathclose{\copy\@brx\kern-0.5\wd\@brx\usebox{\@brx}}}
\makeatother

\usepackage[capitalize]{cleveref}

\begin{document}
	
	\title{Nonadiabatic Renormalization Group for  Strongly Coupled Multiscale Quantum Systems}
\author{Bing Gu}
\email{gubing@westlake.edu.cn}
\affiliation{Department of Chemistry \& Department of Physics, Westlake University, Hangzhou, Zhejiang 310030, China}
\begin{abstract}
Complex quantum systems are often multiscale in nature with strong interactions between  scales. We present a novel idea of iteratively suppressing, rather than tracing out, the fast high-energy degrees of freedom for strongly correlated quantum systems  with multiple energy scales in a non-perturbative way, termed nonadiabatic renormalization group. This leads to a quantum geometric structure of a nested fiber bundle whereby each fiber of a layer is itself a fiber bundle of the next layer. 
The nonadiabatic renormalization group brings a new type of tensor network states that shares physical legs among ``sites'' and encodes quantum entanglement beyond conventional matrix product states. 
We demonstrate how to apply nonadiabatic renormalization group  to different type of problems  including an interacting boson model and  the ab initio quantum chemistry problem with interacting electrons. 
\end{abstract}
\maketitle

\section{Introduction}

Most of the difficult problems in physics, chemistry, and materials science involves many degrees of freedom (DOFs) spaning multiple energy (or time) scales that are strongly coupled. Such problems arise in e.g. condensed matter physics, lattice field theory, open quantum systems, quantum chemistry, vibrational structure, and quantum phase transition. 
As a prominent example, in molecular science, each molecule represents a multiscale quantum system with many interacting electrons strongly coupled to the nuclei. 
Computer modeling of such systems is a significant challenge due to the exponential scaling of required computational resources with system size.  
From a time-dependent pespective, if the degrees of freedom at energy scale $\Lambda$ is excited by external stimutus, then all degrees of freedom associated with energy scales $\mu < \Lambda$ will be excited due to couplings between different scales; whereas the DOFs for higher-energy scales $  \Lambda > \mu$ can be assumed to follow the motion of the slower degrees of freedom.

Renormalization group provides a powerful conceptual framework understanding multiscale systems \cite{kadanoff1966, wilson1983, glazek1993}. The concept of renormalization group has led to the development of powerful numerical techniques for multiscale systems including the numerical renormalization group for quantum impurity models and density matrix renormalization group for one-dimensional lattice systems \cite{bulla2008, white1992, ma2022,cirac2021}.   Such methods often involve a coarse graining step that prevents the exponential growing of computational cost with system size, althought with different criteria. 

Here we present a new kind of renormalization group for multiscale quantum systems, termed nonadiabatic renormalization group (NARG). Conceptually, there are two main  ingrediants to NARG. 
The first  is  an idea of nonadiabatic strong coupling representation to describe the strong interaction between  high- and low-energy degrees of freedom. This representation is applied recursively to a multiscale problem with degrees of freedom ordered by energy accendingly.  The second ingrediant is a coarse graining step by truncating the block Hamiltonian to prevent the exponential grow of the Hilbert space but retaining the relevant configurations.  
NARG iteratively suppress, but not trace out as in an effective action approach, the high-energy degrees of freedom. That is, the high-energy degrees of freedom are not completely ignored but are kept in the basis set so that its observable and correlation functions can still be obtained. The strong interaction between different scales is accounted for by the quantum geometric structure of a nested fiber bundle.

The employing of strong coupling representation with adiabatic-like states leads to a fiber bundle-like structure with nontrivial quantum geometry. The quantum geometry of the projected fiber bundle is usually characterized by the non-Abelian quantum geometric tensor including the Riemmman geometry in the real part and Berry curvature in the imaginary part \cite{provost1980, xie2025c, berry1984}. 
A major part of the inter-scale interaction is built into the adiabatic energies and states, whereas all the residual non-adiabatic effects are encoded in this quantum geometry.   
However, as we have shown in the context of strongly correlated electron-nuclear dynamics \cite{gu2023b,gu2024a,xie2025c}, the challenge is how to describe nontrivial quantum geometry of a (pseudo)fiber bundle as  
the quantum geometric tensor is singular for topologically nontrivial fiber bundles. 
This occur frequenty in strongly coupled systems due to (near)degeneracy. For example,  electronic degeneracy, particularly conical intersections, plays a critical role in the nonadiabatic molecular photochemistry and photophysics \cite{mead1992, wittig2012}. 

 In the strong coupling representation, the quantum geometry is fully accounted for by the singularity-free  global overlap matrix between the eigenstates of the parametric Hamiltonian of the fast degrees of freedom \cite{gu2024a, gu2023b,xie2025, zhu2024}. 
It provides a gauge-indepedent, instead of gauge-covariant, effective description of the slow degrees of freedom, therefore removing the difficult gauge fixing problem\cite{zhu2024}. 

Furthermore, we show that NARG leads to a novel type of tensor network states that physical legs are shared among tensors. We refer to such tensor network states  as leg-tied tensor ansatz (LETTA). As a tensor network state,  the ``sites'' in LETTA does not have to correspond to different energy scales as in NARG. That is, LETTA can be applied to e.g. spin models whereby all sites are equivalent.    
As a generalization of the conventional tensor network states, the LETTA opens up the possibility to go beyond the fundmental  limitation on quantum entanglement. 

For concrete applications, we present numerical implementations of NARG to interacting boson model and to ab initio quantum chemistry. A fast convergence of low-lying many-body eigenstates with respect to the retained states (or bond dimension) is observed for the bosonic model. For the quantum chemistry, it is shown that NARG can capture a large portion of electron correlation in the hydrogen chain with moderate ``bond dimension''.

\section{Strong coupling representation for two-scale problems}
We start by introducing the first critical ingrediant of NARG, that is, a nonadiabatic  decomposition for a strongly coupled bipartitioned system with different energy scales, referred to as strong coupling representation because this decomposition becomes more efficient when the coupling strength grows. 
Consider a generic system with its degrees of freedom bipartitioned into slow and fast degrees of freedom that are strongly coupled. This two-scale problem arises in many problems. For example, in molecular science, the electronic motion is much faster than the nuclear motion; in vibrational-rotational dynamics,   the vibrational motion is much faster than the rotational motion; in attosecond science involving correlated core-valence electron motion,   the energy scale of core electrons is much larger than the valence electron motion.
For such systems described by the Hamiltonian $H = H_\text{f} + H_\text{s} + V(x_\text{f}, x_\text{s})$, 
where $x_\text{f}$ ($x_\text{s}$) denotes fast (slow) degrees of freedom, that can be either discrete or continuous. 
The common treatment is to first diagonalize $H_\text{f}$ and $H_\text{s}$ to obtain the interaction-free states $\phi_n(x_\text{f})$ and $\chi_m(x_\text{s})$. The product basis set $\phi_n(x_\text{f})\otimes \chi_m(x_\text{s})$ is then constructed to diagnalize the full Hamiltonian. This is appropriate for weakly coupled systems. However,  
 such a product basis set is inappropriate to describe a strongly coupled  system. 
When $x_\text{f}$ and $x_\text{s}$ are continuous variables, a more physically motivated ansatz is given by Born and Huang for molecules (i.e. coupled electron-nuclear problems)  \cite{born1988}
$
{\Psi}(x_\text{f}, x_\text{s}) = \sum_{\alpha}  {\phi_\alpha(x_\text{f}; x_\text{s})}  {\chi_\alpha(x_\text{s})}
$ 
where ${\phi_\alpha}(x_\text{f}; x_\text{s})$ is the adiabatic eigenstate of the fast ``electronic'' Hamiltonian that parameterically depend on the slow ``nuclear'' degrees of freedom $H(x_\text{f}; x_\text{s}) = H_\text{f} + V(x_\text{f}; x_\text{s})$, and $\ket{\chi_\alpha} $ is the nuclear wavepacket associated with the adiabatic state $\ket{\phi_\alpha(x_\text{s})}$. 
The adiabatic states are defined by the eigenstates of the electronic Hamiltonian $H(x_\text{f}) = H_\text{f} + V(x_\text{f}; x_\text{s}) + V(x_\text{s})$, where $H_\text{f}$ is the Hamiltonian for the fast degrees of freedom and $V(x_\text{f}; x_\text{s})$ is the fast-slow interaction, assumed to depend on coordinates as in e.g. Coulomb interaction. 
The Born-Huang expansion is better than a direct product basis set because, qualitatively speaking, the adiabatic states already encode a large amount of correlation (entanglement) between the slow and fast degrees of freedom.

However, the challenges with the Born-Huang ansatz is that the coupling between the slow and fast degrees of freedom is due to  the slow kinetic energy operator, referred to as derivative couplings following the terminology in chemical physics.  	
For a non-relativistic dynamical variable $x_\text{s}$, the  kinetic energy operator is $\hat{T}_\text{s} = - \frac{\hbar^2}{2 m_\text{s}}\pd[2]{}{x_\text{s}}$ so that there are four derivative couplings including the first- and second-order nonadiabatic couplings, vector potentials, and the scalar potentials (known as the diagonal Born-Oppenheimer corrections) \cite{mead1979}. 
This requires the adiabatic states $\ket{\phi_\alpha(x_\text{s})}$ to be analytic and single-valued with respect to the slow degrees of freedom. This condition, however, cannot  be satisfied in general. In polyatomic molecules, it is known that  state degenacy (e.g. conical intersections) are ubiquitous \cite{domcke2011}. 
This triggers 
all non-Born-Oppenhaimer couplings  to diverge at electronic degeneracy due to the vanishing of  energy gap \cite{xie2025}.

To address this challenge, we introduce a nonadiabatic decomposition that is a local trivialization of the  fiber bundle. Specifically, for coordinate-dependent couplings $V(x_\text{f}; x_\text{s})$, we employ the eigenstates of the slow interaction operator (i.e. coordinate operator) as the basis set. 
\be 
\set{ \ket{\phi_\alpha(x_\text{s}^n)} \otimes  \ket{x^{n}_\text{s}} } 
	 \label{eq:nonadiabatic_decomposition} 
\ee 
Here $\ket{x_\text{s}^n}$ is the $n$-th eigenstate of the slow coordinate operator $\hat{x}_\text{s}$ in a finite basis representation with eigenvalue $x_\text{s}^n$,  $\ket{\phi_\alpha(x_\text{s}^n)}$ is the ``adiabatic'' (or conditional) eigenstate at the $n$-th configuration of the slow degrees of freedom (i.e., eigenstates of the fast Hamiltonian $H_\text{f}(x_\text{s}^n) = \hat{T}_\text{f} + V(x_\text{f}; x_\text{s}^n) + V_\text{s}(x_\text{s}^n)$ that are parameterized by the slow coordinates with corresponding eigenvalues $V_{n\alpha}$), and $C_{n \alpha}$ is the expansion coefficient. 
The total Hamiltonian matrix in this basis set \cref{eq:nonadiabatic_decomposition} reads 
\be 
H_{m\beta, n\alpha} = T_{mn}A_{m\beta, n\alpha} + V_{n\alpha} \delta_{mn} \delta_{\beta \alpha}
\ee 
where $T_{mn}$ is the slow kinetic energy operator matrix in the basis set $\ket{x_\text{s}^n}$. 
Here  
\be A_{m\beta, n\alpha} = \braket{\phi_\beta(x_\text{s}^m) | \phi_\alpha(x_\text{s}^n)} \label{eq:overlap_matrix} \ee	 
is the \emph{global} overlap matrix between the eigenstates $\ket{\phi_\beta(x_\text{s}^n)}$ and $\ket{\phi_\alpha(x_\text{s}^m)}$ of the high-energy degrees of freedom at different configurations of the slow degrees of freedom. Its modulus squared $\abs{A_{m\beta, n\alpha}}^2$ are known as the fidelity in quantum information theory, which is a measure of the ``closeness'' between two states. The phase of $A_{m\beta, n\alpha}$ acccounts for the geometric phase effect. 
We call such as a decomposition strong coupling representation as the basis set becomes the exact eigenstates of the total system in the infinity  coupling strength limit. 

The global overlap matrix is the key ingredient of the strong coupling representation. It fully accounts for the quantum geometry of the  fiber bundle of ``adiabatic'' eigenstates and the strong coupling between the slow and fast degrees of freedom. 
%
%
In the language of differential geometry, the manifold of adiabatic states forms a fiber bundle with the configuration space as the base space and the adiabatic states at each configuration as the fiber. The quantum geometry of the fiber bundle is characterized by the non-Abelian quantum geometric tensor $Q_{\mu \nu}^{\beta \alpha}(x_\text{s}) = \braket{\pa_{\mu} \phi_\beta(x_\text{s}) | 1 - \hat{\mc{P}}(x_\text{s})| \pa_{\nu} \phi_\alpha(x_\text{s})}$ \cite{xie2025c}, consisting of the Riemann quantum metric ($\Re Q$) and Berry curvature ($\Im Q$) \cite{provost1980}.  
This fiber bundle is always geometrically nontrivial due to inevitable state truncation (or regularization),  meaning it is impossible to find a global trivialization due to the dependence of the adiabatic states on the slow degrees of freedom. Note that it does not require state degeneracy to have a nontrivial quantum geometry, although degeneracy, either symmetry-protected or accidental, often enriches it \cite{bersuker2006}. The exponential convergence of the truncated states is crucial for this representation to be practically useful. Without truncation, the (infinity) adiabatic states are merely a unitary transformation of an arbitrary complete basis set in the fast degrees of freedom. 
The strong coupling representation is in essence a discretized local trivialization of the fiber bundle consisting of the adiabatic eigenstates as the fiber and the slow-operator eigenvalues as the parameter space. 
%

The  strong coupling representation can be straightforwardly applied for discrete variables, e.g., spin and fermion variables, although there may not necessarily be a continuous parameter space to define the  quantum geometric tensor. 

If the fast-slow interaction involves non-commutating operators (e.g. $V = A \otimes x_\text{s} + B\otimes p_\text{s}$),  it is then not possible to find common eigenstates of all interaction operators.  There are several choices one can choose to deal with this adiabatic flustration. First, it may be possible to remove the adiabatic flustration by a similarity transformation. Secondly, when there is a dominate interaction, we can still choose the eigenstates of that particular  term.  Another possibility is to choose states that minimize the uncertainty of the interaction operators. 


\section{From two-scale to many-scale}
 We now generalize the above idea of the strong coupling representation from two-scale to multiscale systems, whereby the degrees of freedom are separated by energy scales in descending order $x_i, i=0, \cdots, L-1$. Each $x_i$ can contain multiple degrees of freedom that are in the same energy scale. 
Using molecules as an example, each molecule is an intrinsic multiscale system. Besides the separation of electronic and nuclear motion underlying the Born-Oppenhaimer approximation\cite{born1927}, the electronic motion can be further separated into the core and valence electrons, which can differ in energy scale by a factor of 1000 (keV vs. eV). In turn, the nuclear motion can be separated into vibrational and rotational motions. 
Moreover, the vibrational motion can be even further separated into high-frequency and low-frequency modes. Hence,  there is a hierarchy of interactions between different energy scales: the core electrons are strongly coupled to the valence electrons, the valence electrons are coupled to the vibrations, the vibrational motion is strongly coupled to the rotational motions.

The main idea of the nonadiabatic renormalization group (NARG) is as follows. For illustration, we consider continuous variables with coordinate-dependent couplings. We start by grouping and ordering the degrees of freedom by their energy scale $x_i, i=0, \cdots, L-1$.

(i) Consider first the two highest energy scales $x_0, x_1$, we choose a primitive basis set for $x_1$, i.e., $\ket{x^{[1] n}}, n = 0, 1, \dots, N_1-1$, $N_j$ is the number of basis sets for each DOF. 
For coordinate couplings, the eigenstates in a finite basis representation is the so-called discrete variable representation basis set\cite{light2000}. 
We construct the adiabatic eigenstates of fast scale ($x_0$) that parameterically depends on $x_1$ 
\be
\del{H_0 + V(x_0; x_1^{[j_1]})} \ket{\phi_{j_1 \alpha_0}}  = E_\alpha(x_1^{[j_1]}) \ket{\phi_{j_1 \alpha_0}}
\ee 
where $\braket{x_0| \phi_{j_1 \alpha_0}} = \phi_{\alpha_0}(x_0; x_1^{[j_1]})$. Here we include the potential energy operator for the slow degrees of freedom in $V(x_0, x_1)$. 

To solve the total Hamiltonian for both DOFs, we include the kinetic energy operator of $x_1$.  The kinetic energy operator matrix elements can be analytically calculated in most discrete variable representation basis sets. 
With the basis sets for the composite system
\be
\ket{j_1 \alpha_0} \equiv \ket{\phi_{j_1 \alpha_1}} \otimes \ket{x^{[1] {j_1}}}, 
\label{eq:113}
\ee 
where $j_1 = 0, 1, \cdots, N_1-1, \alpha_1 = 0, 1, \cdots, D-1$, 
the total Hamiltonian matrix reads 
\begin{equation}
	\bf H = \braket{j_1' \alpha_0' | H(x_0; x_1) + \hat{T}_1 | j_1 \alpha_0 } =  {T}^{[1]}_{j_1' j_1} A_{j_1' \alpha_0', j_1 \alpha_0} + \bf V  
\end{equation}
and $\bf V$ is the potential energy matrix. The overlap matrix $A_{j_1'\alpha_0', j_1\alpha_0} = \braket{\phi_{j_1' \alpha_0'} | \phi_{ j_1 \alpha_0}}$ accounts for the quantum geometry of the manifold of the adiabatic eigenstates. It includes both the inflences of the quantum metric and geometric phase effects. 
A truncation of the adiabatic eigenstates are required to prevent the exponential growth of the Hilbert space, that is, the coarse-graining step. Retaining $D$ states  
\be 
\begin{split}
\phi_{\alpha_1} (x_{0}, x_1) &= \sum_{\alpha_0, j_1} C^{\alpha_1}_{\alpha_0 j_1} \ket{\phi_{j_1 \alpha_0}} \otimes \ket{x^{[1] {j_1}}}
\label{eq:111}
\end{split}
\ee
where $\alpha_1 = 0, 1, \cdots, D-1$.



(ii) We now add the $x_2$ DOFs of energy scale lower than $x_0$ and $x_1$; it may couple to the fast degrees of freedom through $V( x_1, x_2)$. Similarly as in the last step, we choose a DVR basis set for $x_2,  \ket{x_2^{j_2}}, j_2 = 0,1,  \cdots, d_2-1$ then we construct adiabatic eigenstates of $x_0, x_1$ as a function of $x_2$ using the eigenstates $\ket{\phi_{\alpha_1}}$ as a basis set. 
 Specially, for each $x_2^{j_2}$, we rebuild the Hamiltonian matrix for the $x_0, x_1$ degrees of freedom taking into account the interaction between $x_1$ and $x_2$ . 

\be
\del{ H(x_0, x_1) + V(x_0, x_1; x_2^{n_2}) } \ket{\phi_{j_2 \alpha_2}} = E_{j_2 \alpha_2}   \ket{\phi_{j_2 \alpha_2}} 
\ee 
That is, the adiabatic eigenstate of $x_0, x_1$ as a function of $x_2$ reads 
\be 
\ket{\phi_{j_2 \alpha_1}} = \sum_{j_1 \alpha_1} C^{j_2 \alpha_1}_{j_1 \alpha_0} \ket{\phi_{j_1\alpha_0}} 
\label{eq:116}
\ee

(iii)  Repeat step 2 until degrees of freedom of all scales are included. 

\begin{figure}[h]
\centering
\includegraphics[width=0.45\textwidth]{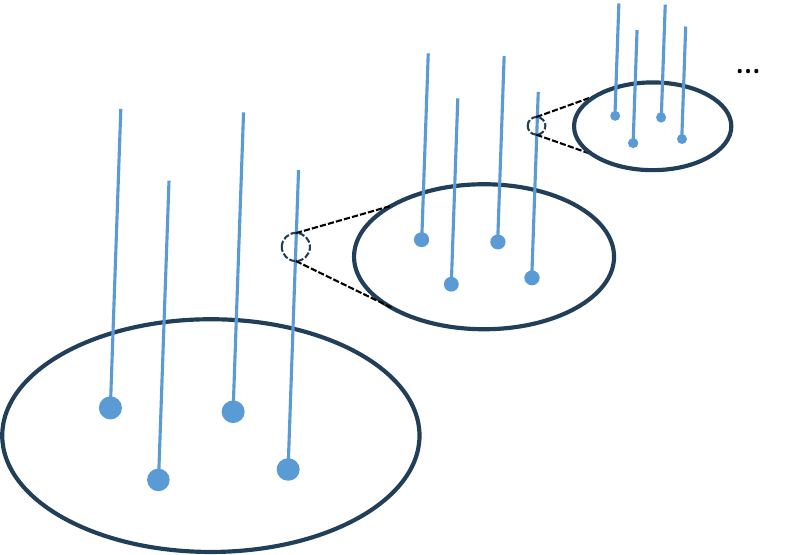}
\caption{Schematic of NARG as a nested fiber bundle whereby the fiber of each layer is itself a fiber bundle.}
\label{fig:fiber}
\end{figure}

The whole proceture leads to a nested (or continued) fiber bundle structure, schematically shown in \fig{fig:fiber}. It starts with the adiabatic states of $x_0$ as the fiber and $x_1$ as the parameter space. Upon including $x_2$, it acts as the new parameter space whereas the fiber becomes the adiabatic states of $x_0$ and $x_1$ and this structure continues until all scales are included.

\textit{Beyond nearest-neighbor scale - }
We can further generalize the above procedure to account for strong interaction between non-nearest-neighbours. 
Consider the case in which when we add the second scale degrees of freedom $x_2$ to the system, it not only strongly couples to the nearest-neighbor scale $x_1$ but also to the next-nearest-neighbor $x_0$, then we can calculate the adiabatic eigenstates for both $x_0$ and $x_1$. That is, for each $x_2^{j_2}$, we rebuild the Hamiltonian matrix for the $x_0, x_1$ degrees of freedom taking into account the interaction between $x_0$ and $x_2$. For each $x^{[2] j_2}$, the adiabatic eigenstates for $x_0$ are obtained by diagonalizing  the Hamiltonian $H(x_0; x_1^{[j_1]}, x_2^{[j_2]})$.
For each $j_1$, we obtain $D$ adiabatic eigenstates for $x_0$ with lowest energy, $\ket{\phi^{\alpha_1}_{j_1j_2}} = C^{\alpha_1}_{j_0 j_1j_2} \ket{x^{[0] j_0}}$. With this set of adiabatic states, we can construct the Hamiltonian matrix $H(x_0, x_1; x_2^{j_2})$ for the $x_0, x_1$ degrees of freedom at a fixed $x_2$. Scanning through all $x_2^{j_2}, j_2 = 0, 1, \cdots, d_2-1$, then we can build the Hamiltonian matrix for $x_0, x_1, x_2$.

\section{A new type of tensor network states}

\begin{figure}[h]
\centering
\includegraphics[width=0.4\textwidth]{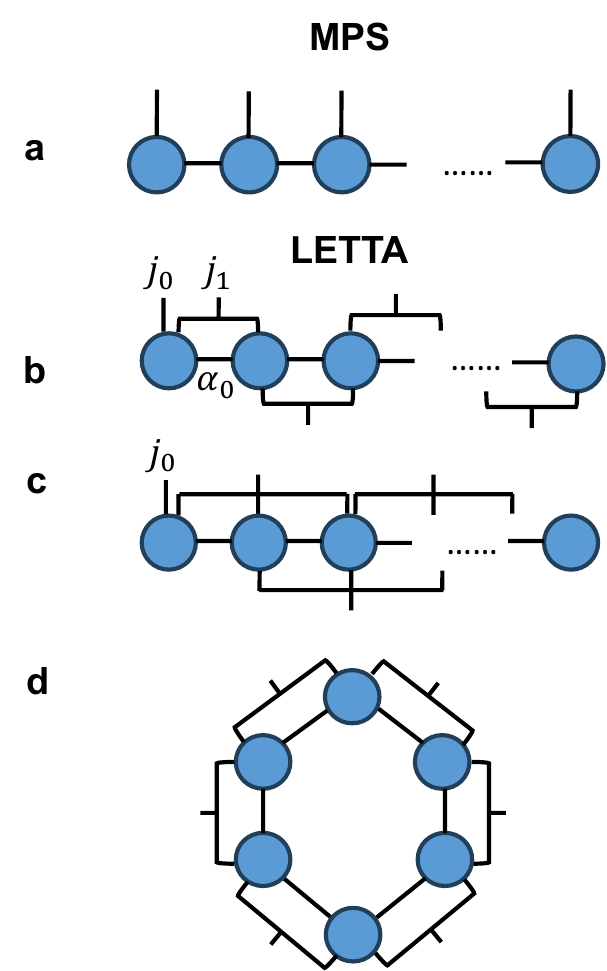}
\caption{Schematic of the leg-tied tensor ansatz (LETTA) in comparison with the conventional matrix product states (MPS). Physical legs can be shared between (a) nearest neighbours and  (b) next-nearest-neighbours. (d) A LETTA with peridic boundary condition.}
\label{fig:letta}
\end{figure}

We now show that, interestingly, the successive iteration in NARG leads to a new type of tensor network representation of many-body states. To do so, inserting \eq{eq:113} into \eq{eq:111} yields 
\be  
\phi_{\alpha_1} (x_{0}, x_1)  = 
{C^{j_0 j_1}_{\alpha_0}} C_{\alpha_0 \alpha_1}^{j_1}.   \ket{x^{[0] {j_0}}}  \otimes \ket{x^{[1] {j_1}}} 
\label{eq:114}
\ee 
where we have reshaped the expansion coefficients.
We can further insert \eq{eq:114} into \eq{eq:116} leads to a matrix product-like state 
\be 
\ket{\Psi^{\alpha_2}_{j_0 j_1 j_2 }} = \sum_{\alpha_0 \alpha_1 \alpha_2} C^{j_0 j_1}_{\alpha_0} C^{j_1 j_2}_{\alpha_0 \alpha_1} C^{j_1 j_2}_{\alpha_1 \alpha_2}  \ket{x^{[0]j_0}}\otimes \ket{x^{[1] {j_1}}} \otimes \ket{x^{[2] {j_2}}} 
\ee 
When all scales are included, the NARG leads to a tensor network-like states in which multiple tensors share physical legs,
\be 
\begin{split}
\ket{\Psi} =& \sum_{\alpha_0 \alpha_1 \dots } C^{j_0 j_1}_{\alpha_0} C^{j_1 j_2}_{\alpha_0 \alpha_1} \cdots C_{\alpha_{L-2} \alpha_{L-1}}^{j_{L-1} j_L} \\ 
&  \ket{x^{[0]j_0}}\otimes \ket{x^{[1] {j_1}}} \otimes \ket{x^{[2] {j_2}}} \otimes \cdots \otimes \ket{x^{[L] {j_L}}}
\end{split}.
\ee 
We call such states  LEg-Tied Tensor Ansatz (LETTA). In it, the entanglement between different energy scales (or ``sites'') is encoded in both the basis set and the coefficients. 
Different from the conventional matrix product states with one physical leg and two virtual legs, each tensor in LETTA has four physical legs. 
In conventional matrix product state  
\be
\Psi_\text{MPS} = \sum_{\alpha_i=1}^D M^{[1] \sigma_1}_{\alpha_0 \alpha_{1}} M^{[2] \sigma_2}_{\alpha_1 \alpha_{2}} \cdots M^{[L] \sigma_L}_{\alpha_{L-1} \alpha_L} \ket{\bm \sigma}, 
\ee  
each tensor has only one physical leg and two virtual legs. 
A graphical representation of LETTA is shown in \cref{fig:letta}b, in comparison to the conventional MPS (\cref{fig:letta}a). With the tensor diagram, it is straightforward to envision a LETTA with long-range leg sharing (e.g, with next-nearest-neighbor leg-sharing \cref{fig:letta}c) and also with periodic boundary conditions, \cref{fig:letta}d. 

The critical concept for LETTA is nonadiabaticity whereas for conventional matrix product states is entanglement. 
These two concepts are highly related to each other. However, an adiabatic state with vanishing nonadiabaticity (e.g. the  Born-Oppenheimer state for molecules) can have enormous entanglement encoded in the adiabatic following. Nonadiabaticity is like condtional entropy, measuring the uncertainty of the high-energy degrees of freedom given the state of the slow degrees of freedom. In the adiabatic limit, the high-energy degrees of freedom are completely determined by the slow degrees of freedom.  For example, the entangled photon pair created by the parametric down conversion can be described by a joint spectral amplitude that is roughly $J(\omega_s, \omega_i) \propto \delta(\omega_s + \omega_i - \omega_\text{p})$ due to energy conservation. The two photons are highly entangled in the frequency domain, but fully ``adiabatic''. Once the frequency of one photon is fixed, the twin photon frequency is completely determined.

The LETTA with couplings across two scales (i.e., next-nearest-neighbor) is schmeticially shown in \cref{fig:letta}c. Comparing to the nearest-neighbor case whereby two tensors share a leg, here three tensors share a physical leg. Although physically it is expected that degrees of freedom across many scales cannot be strongly coupled as energy cannot be easily exchanged between different scales (e.g., electronic motion can be strongly coupled to vibrational motion, but only weakly to the molecular rotation), we can immediately generalize NARG and LETTA to couplings across $N$ scales, where a physical leg is shared among $N$ tensors.

%


With LETTA, we can relax the constraint in NARG that each ``site''  has to correspond to an energy scale. In other words, LETTA can be applied to systems where sites are not associated with energy scales as in e.g. spin and lattice models. Optimization algorithms analogous to density-matrix renormalization group and variational matrix product states can be developed  for LETTA.  LETTA   goes beyond NARG in, loosely speaking, the same way as tensor network states goes beyond the numerical renormalization group.   It also opens up the possibility to create mixed tensor network states with both MPS and LETTA.

\section{Applications}


\subsection{Interacting bosons}
We first apply the NARG to an interacting boson model 
\be  
H = \sum_i  \frac{ \omega_i}{2} \qty( p_i^2 + x_i^2 + \lambda_i x_i^4) + \sum_{i<j} g_{ij} \sqrt{\omega_i \omega_j} {x_i^2 x_j^2}
\ee 
where the dimensionless parameters $\lambda_i$ and $g_{ij}$ measures the anharmonicity and the mode-coupling strength, respectively. $\omega_n, n =1, 2, \cdots, N$ are the  frequencies. Such model can represent e.g. the vibrational eigenstates problem with strong anharmonicity 
A brute force exact diagonalization of the Hamiltonian in a direct product basis set is not feasible as the computational cost increases exponentially with the number of modes.

\begin{figure}[h]
	\centering
	\includegraphics[width=0.46\textwidth]{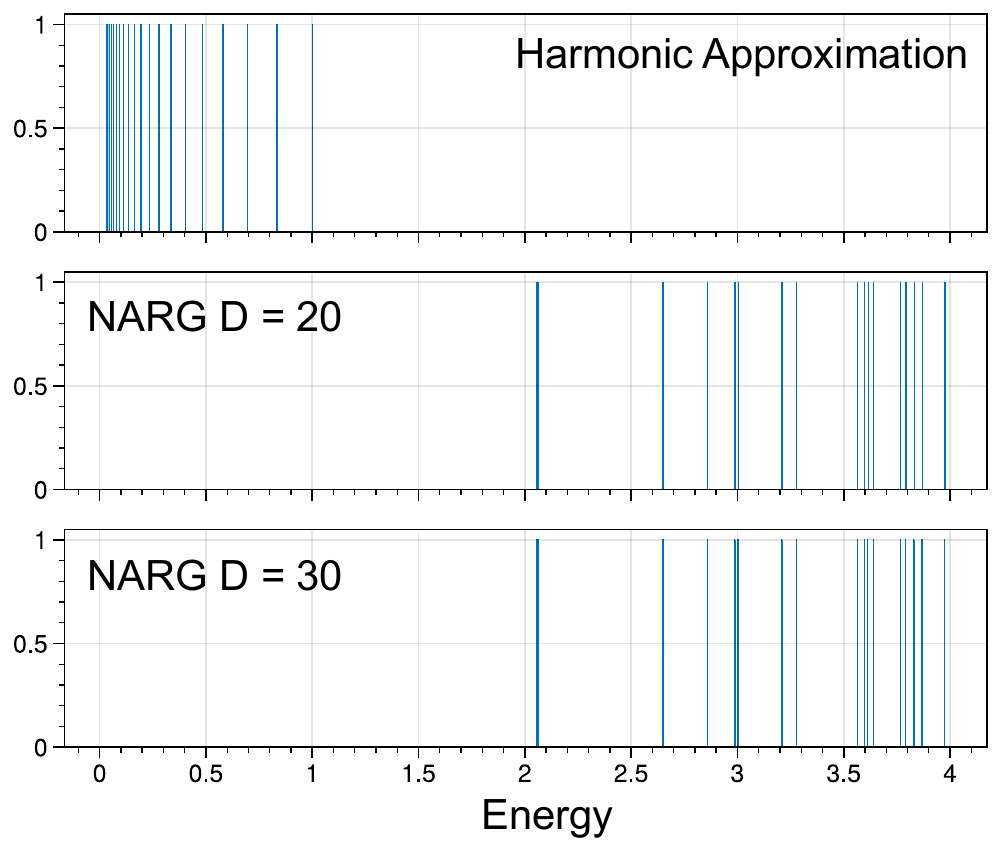}
	\caption{Low-lying eigenstates of an strongly interacting boson model with 20 modes. The lowerst 16 eigenvalues are shown, with a rapid convergence with respect to the number of retained adiabatic states $D$. }
	\label{fig:vib}
\end{figure}

The 16 lowest eigenstates corresponding to an strongly interacting parameter regime with 20 modes are shown in \fig{fig:vib}.  The harmonic approximation breaks down in this regime, the NARG shows a rapid convergence with respect to the number of retained adiabatic states $D$. The computation takes only less than 20 seconds with a purely Python implementation.

\subsection{Ab initio Quantum Chemistry}

We now demonstrate how NARG can be used to describe interacting fermions, specifically,  electron correlation in ab initio quantum chemistry. 
The ab initio electronic Hamiltonian in second quantization is given by 
\be 
H = \sum_{i,j} \sum_{\sigma = \uparrow, \downarrow}  t_{ij} c_{i\sigma}^\dag c_{j\sigma} + \frac{1}{2} \sum_{i,j,k,l} \sum_{\sigma, \tau} v_{ijkl} c_{i\sigma}^\dag  c^\dag_{k \tau} c_{l \tau} c_{j\sigma} 
\ee 
where $ 
t_{ij} = \braket{i |  -\half \grad^2 + V(\bf r) | j} 
$ is the core matrix elements consisting of the electronic kinetic energy operator and   the electron-nuclear Coulomb attraction $V(\bf r)$, and $v_{ijkl}$ is the two-electron Coulomb repulsion integrals. 
We use caonical molecular orbitals here as the one-electron basis set as defined in the Hartree-Fock theory. An immediate question is that how should we order the molecular orbitals.  
In the quantum chemistry NARG, we order the molecular orbitals in descending order of energy, i.e.,  core orbitals first and then valence orbitals because the valence orbitals should follow the motion of core orbitals.

\begin{figure}[hbpt]
	\centering
	\includegraphics[width=0.46\textwidth]{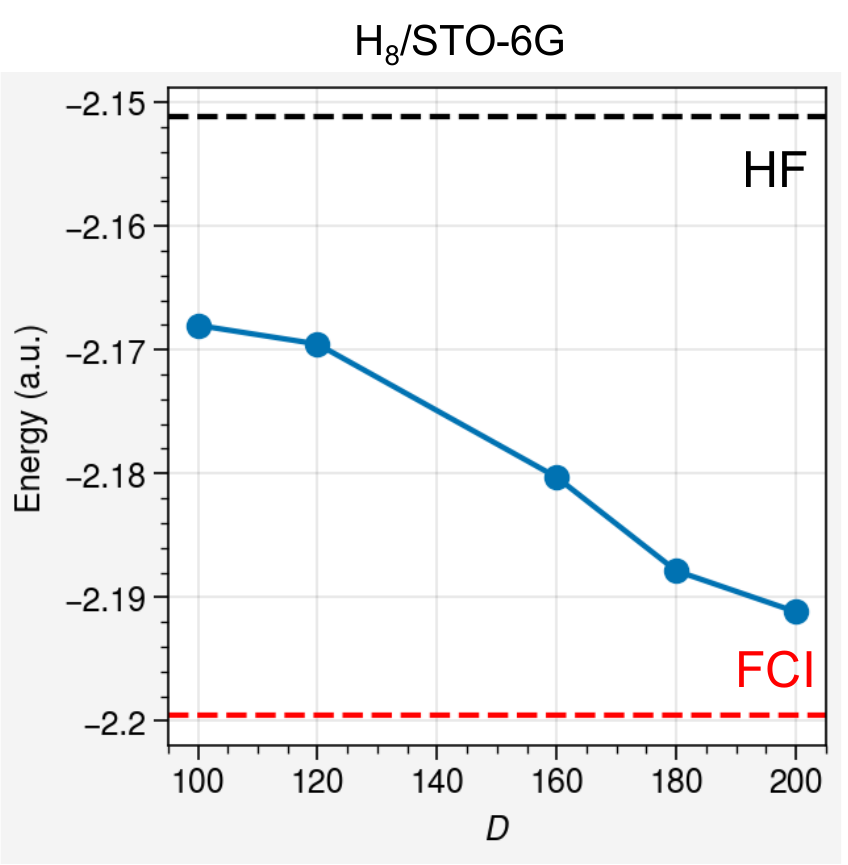}
	\caption{NARG for ab initio quantum chemistry. Electronic correlation is gradually incorporated into the NARG by increasing the retained adiabatic states. More than 80\% of the correlation energy in the hydrogen chain H$_8$ is captured by $D=200$.}
	\label{fig:h8}
\end{figure}


 We briefly outline the key steps in the quantum chemistry  NARG. A detailed description of the method and of the implementation  is in Sec. S2. 
The computation is initialized by a mean-field (e.g. Hartree-Fock) calculation to obtain the molecular orbitals and matrix elements.   It is possible to define an active space such that only a subset of molecular orbitals are considered in the following steps.  
We label the  molecular orbitals in descending order of energy as $1, 2, \cdots, L$ so that the core orbitals is considered first and then valence orbitals.
We start by a full configuration interaction (i.e., exact diagonalization) of the first $\ell$ spin-orbitals,  truncated to $D \le d^\ell$ states by energy. 
Now add the $\ell + 1$ molecular orbital with $d = 4$ states. For each state $\ket{\sigma^{\ell+1}} \in \set{\ket{0}, \ket{\uparrow}, \ket{\downarrow}, \ket{\uparrow \downarrow}}$, we rebuild the Hamiltonian for the first $\ell$ orbitals, $H[\sigma^{\ell+1}]$, including the interaction between the block and the $\ell+1$th orbital,  diagonalize it  to obtain a new set of ``adiabatic'' eigenstates $\ket{\phi_\alpha(\sigma^{\ell+1})}$. 
With this set of  adiabatic eigenstates $ D \times d$, we can build the composite superblock Hamiltonian for the total $\ell+1$ orbitals followed by diagonalization. 
We repeat this process for the $\ell+2$ orbital until we reach the total number of orbitals, $L$ in the active space. The Jordan-Wigner transformation is used to ensure the anticommutation relations in the fermion-to-spin mapping.

The ground state energy of a hydrogen chain  H$_8$ is shown in \fig{fig:h8} using a minimal basis set STO-6G to compare with the full configuration interaction. When increasing the number of retained states, the electronic correlation is gradually captured.  More than 80\% of the correlation energy  is included by $D=200$.


\section{Conclusions}

NARG provides a novel strategy for tackling complex  quantum systems with a hierarchy of strongly coupled energy (or time) scales.  
With an interacting boson model and ab intio quantum chemistry, we have demonstrated how NARG can be exploited to tackle strongly coupled multiscale quantum systems. By this, we can easily envision applications of NARG   to a wide variety of multiscale strongly interacting systems across chemistry, physics, and materials science, including interacting bosons, interacting fermions, lattice field theory, open quantum systems, quantum electrodynamics, Floquet engineering, strong light-matter interaction, quantum molecular dynamics, vibrational dynamics, electron-nuclear dynamics. 
We can moreover forsee how  NARG can be used to study not only the low-energy eigenvalue problems but also non-equilibrium real-time quantum dynamics in a similar way how conventional tensor network states can be used to describe real-time problems. 

 Moreover, NARG suggests a new type of tensor network states, LETTA, with tensors sharing  physical legs for quantum many-body problems. 
By tying the legs of the tensors, LETTA  encodes more entanglement than the conventional tensor network states, and opens up the possibility to go beyond the fundamental limitation of conventional matrix product states. It is intriguing to explore the possibility of applying  LETTA to study strongly correlated many-body systems directly using variational methods. This requires further development of optimization algorithms for such states. 

\section{Acknowledgments}

This work is supported by the National Natural Science Foundation of China (Grant Nos. 22473090 and 92356310).

\appendix
\bibliography{../qchem,../dynamics,../topology,../RG,../tensor}
\end{document}